\documentclass{birkjour}
\usepackage[noadjust]{cite}

\usepackage[utf8]{inputenc}
\usepackage[T1]{fontenc}
\usepackage{newunicodechar}
\usepackage[unicode]{hyperref}
\usepackage{todonotes}
\usepackage{csquotes}
\usepackage{cleveref}
\usepackage{MnSymbol}
\usepackage{mathtools}
\usepackage{tikz-cd}
\usepackage{xspace}
\usepackage{orcidlink}
\usepackage{accsupp}
\usepackage{threeparttable}

\usepackage{currfile-abspath}
\getabspath{\jobname.log}
\ifthenelse{\equal{\theabsdir}{\thepwd}}
    {}{\PassOptionsToPackage{outputdir=\theabsdir}{minted}}

\usepackage{main-pygments-output}
\usepackage[frozencache,cachedir=.]{minted}

\newcommand{\negphantom}[1]{\settowidth{\dimen0}{#1}\hspace*{-\dimen0}}

\newcommand{\mathlib}{\textsf{mathlib}\xspace}

\input{glyphtounicode}
\newunicodechar{∗}{\BeginAccSupp{method=hex,unicode,ActualText=2217}\ensuremath{\ast}\EndAccSupp{}}
\newunicodechar{≠}{\BeginAccSupp{method=hex,unicode,ActualText=2260}\ensuremath{\ne}\EndAccSupp{}}
\newunicodechar{−}{\BeginAccSupp{method=hex,unicode,ActualText=2212}\ensuremath{-}\EndAccSupp{}}
\newunicodechar{Π}{\BeginAccSupp{method=hex,unicode,ActualText=03A0}\ensuremath{\mathrm{\Pi}}\EndAccSupp{}}
\newunicodechar{⨁}{\BeginAccSupp{method=hex,unicode,ActualText=2A01}\ensuremath{\bigoplus}\EndAccSupp{}}
\newunicodechar{∑}{\BeginAccSupp{method=hex,unicode,ActualText=2211}\ensuremath{\mathrm{\Sigma}}\EndAccSupp{}}
\newunicodechar{⦃}{\BeginAccSupp{method=hex,unicode,ActualText=2983}\ensuremath{\{\!|}\EndAccSupp{}}
\newunicodechar{⦄}{\BeginAccSupp{method=hex,unicode,ActualText=2984}\ensuremath{|\!\}}\EndAccSupp{}}
\newunicodechar{≃}{\BeginAccSupp{method=hex,unicode,ActualText=2243}\ensuremath{\simeq}\EndAccSupp{}}
\newunicodechar{∃}{\BeginAccSupp{method=hex,unicode,ActualText=2203}\ensuremath{\exists}\EndAccSupp{}}
\newunicodechar{∀}{\BeginAccSupp{method=hex,unicode,ActualText=2200}\ensuremath{\forall}\EndAccSupp{}}
\newunicodechar{⅟}{\BeginAccSupp{method=hex,unicode,ActualText=215F}\ensuremath{{}^1\!/}\EndAccSupp{}}
\newunicodechar{∞}{\BeginAccSupp{method=hex,unicode,ActualText=221E}\ensuremath{\infty}\EndAccSupp{}}
\newunicodechar{α}{\BeginAccSupp{method=hex,unicode,ActualText=03B1}\ensuremath{\alpha}\EndAccSupp{}}
\newunicodechar{β}{\BeginAccSupp{method=hex,unicode,ActualText=03B2}\ensuremath{\beta}\EndAccSupp{}}
\newunicodechar{ℝ}{\BeginAccSupp{method=hex,unicode,ActualText=211D}\ensuremath{\mathbb{R}}\EndAccSupp{}}
\newunicodechar{ℕ}{\BeginAccSupp{method=hex,unicode,ActualText=2115}\ensuremath{\mathbb{N}}\EndAccSupp{}}
\newunicodechar{ℂ}{\BeginAccSupp{method=hex,unicode,ActualText=2102}\ensuremath{\mathbb{C}}\EndAccSupp{}}
\newunicodechar{λ}{\BeginAccSupp{method=hex,unicode,ActualText=03BB}\ensuremath{\lambda}\EndAccSupp{}}
\newunicodechar{ι}{\BeginAccSupp{method=hex,unicode,ActualText=03B9}\ensuremath{\iota}\EndAccSupp{}}
\newunicodechar{∅}{\BeginAccSupp{method=hex,unicode,ActualText=2205}\ensuremath{\varnothing}\EndAccSupp{}}
\newunicodechar{∪}{\BeginAccSupp{method=hex,unicode,ActualText=222A}\ensuremath{\cup}\EndAccSupp{}}
\newunicodechar{∩}{\BeginAccSupp{method=hex,unicode,ActualText=2229}\ensuremath{\cap}\EndAccSupp{}}
\newunicodechar{∈}{\BeginAccSupp{method=hex,unicode,ActualText=2208}\ensuremath{\in}\EndAccSupp{}}
\newunicodechar{Δ}{\BeginAccSupp{method=hex,unicode,ActualText=0394}\ensuremath{\Delta}\EndAccSupp{}}
\newunicodechar{ᵒ}{\BeginAccSupp{method=hex,unicode,ActualText=1D52}\ensuremath{{}^o}\EndAccSupp{}}
\newunicodechar{₀}{\BeginAccSupp{method=hex,unicode,ActualText=2080}\ensuremath{{}_o}\EndAccSupp{}}
\newunicodechar{ᵖ}{\BeginAccSupp{method=hex,unicode,ActualText=1D56}\ensuremath{{}^p}\EndAccSupp{}}
\newunicodechar{ᵈ}{\BeginAccSupp{method=hex,unicode,ActualText=1D48}\ensuremath{{}^d}\EndAccSupp{}}
\newunicodechar{ₙ}{\BeginAccSupp{method=hex,unicode,ActualText=2099}\ensuremath{{}_n}\EndAccSupp{}}
\newunicodechar{₁}{\BeginAccSupp{method=hex,unicode,ActualText=2081}\ensuremath{{}_1}\EndAccSupp{}}
\newunicodechar{₂}{\BeginAccSupp{method=hex,unicode,ActualText=2082}\texorpdfstring{\ensuremath{{}_2}}{₂}\EndAccSupp{}}
\newunicodechar{₃}{\BeginAccSupp{method=hex,unicode,ActualText=2083}\ensuremath{{}_3}\EndAccSupp{}}
\newunicodechar{ₗ}{\BeginAccSupp{method=hex,unicode,ActualText=2097}\ensuremath{{}_l}\EndAccSupp{}}
\newunicodechar{ₐ}{\BeginAccSupp{method=hex,unicode,ActualText=2090}\ensuremath{{}_a}\EndAccSupp{}}
\newunicodechar{⋏}{\BeginAccSupp{method=hex,unicode,ActualText=22CF}\ensuremath{\curlywedge}\EndAccSupp{}}
\newunicodechar{↥}{\BeginAccSupp{method=hex,unicode,ActualText=21A5}\ensuremath{\upmapsto}\EndAccSupp{}}
\newunicodechar{̅}{\BeginAccSupp{method=hex,unicode,ActualText=0305}\ensuremath{\overline{\phantom{x}}\negphantom{x}}\EndAccSupp{}}
\newunicodechar{▸}{\BeginAccSupp{method=hex,unicode,ActualText=25B8}\ensuremath{\blacktriangleright}\EndAccSupp{}}
\newunicodechar{∥}{\BeginAccSupp{method=hex,unicode,ActualText=2225}\ensuremath{\|}\EndAccSupp{}}
\newunicodechar{⊗}{\BeginAccSupp{method=hex,unicode,ActualText=2297}\ensuremath{\otimes}\EndAccSupp{}}

\setminted{
  fontsize=\scriptsize,
  frame=single,
  breaklines,
  samepage
}

 \theoremstyle{definition}
 
 \theoremstyle{remark}

 \numberwithin{equation}{section}

\pdfinterwordspaceon

\newcommand{\ed}{\end{document}}
\begin{document}

\title{Formalizing Geometric Algebra in Lean}

\author{Eric Wieser\,\orcidlink{0000-0003-0412-4978}}
\address{Cambridge University Engineering Department, Cambridge, UK}
\email{efw27@cam.ac.uk}

\author{Utensil Song\,\orcidlink{0000-0003-3994-4466}}
\address{Independent, Shenzhen, China}
\email{utensilcandel@gmail.com}


\date{\today}

\begin{abstract}

This paper explores formalizing Geometric (or Clifford) algebras into the Lean 3 theorem prover, building upon the substantial body of work that is the Lean mathematics library, \mathlib.
As we use Lean source code to demonstrate many of our ideas, we include a brief introduction to the Lean language targeted at a reader with no prior experience with Lean or theorem provers in general.

We formalize the multivectors as the quotient of the tensor algebra by a suitable relation, which provides the ring structure automatically, then go on to establish the universal property of the Clifford algebra.
We show that this is quite different to the approach taken by existing formalizations of Geometric algebra in other theorem provers; most notably, our approach does not require a choice of basis.

We go on to show how operations and structure such as involutions, versors, and the $\mathbb{Z}_2$-grading can be defined using the universal property alone, and how to recover an induction principle from the universal property suitable for proving statements about these definitions.
We outline the steps needed to formalize the wedge product and $\mathbb{N}$-grading, and some of the gaps in \mathlib that currently make this challenging.

\end{abstract}

\maketitle

\section{Introduction}
\label{sec:intro}

Geometric Algebra provides a language for manipulating geometry algebraically, which unifies many geometric applications of complex numbers, quaternions, and linear algebra into a single mathematical language.

There are software packages in various languages today which support working with and developing intuition for Geometric Algebra.
Most of these packages are designed for numerical computation, typically targeted somewhere along the spectrum with ease of use at one end and efficiency at the other.
However, it is the packages designed for symbolic manipulation which are most relevant to the topic of this paper, such as the packages for Python \cite{galgebra}, Maple \cite{maple-bigebra}, and Mathematica \cite{mathematica-clifford}.
These packages are built on top of computer algebra systems (CAS), and typically consist of definitions of the objects and operators of GA, and sometimes some specialized routines to simplify expressions.

A key feature in a CAS is the ability to simplify expressions to save the user from doing so manually, which can be slow and error-prone.
Such a system is preloaded with a set of rules, such as $x(y + z) = xy + xz$ which can be used to manipulate expressions.
These rules are in a sense axiomatic; a CAS does not provide proofs of these rules, and so trust has to be placed in both the author of the rules, and in the procedures used to apply them.

Theorem provers take quite a different approach---the base set of axioms is very small (for example, those of Zermelo–Fraenkel set theory), and every other \enquote{rewrite rule} must be proven from these.
This approach drastically reduces the amount of trust needed from the user---instead of having to place trust in every single rewrite rule, their trust can be concentrated onto: the theorem prover itself, its choice of axioms, and the fact that the definitions used in the rewrite rule correspond to their established meaning\footnote{e.g. a rule can state $1 \times 1 = 2$ if it defines $\times$ as $+$}.

One such theorem prover is Lean \cite{lean_2015}, developed by Microsoft Research.
We choose to use this theorem prover for our formalization not because we believe it is objectively better, but because of the rapidly growing library and community forming around it, which we describe more in \cref{sec:lean_mathlib}.

This paper will start in \cref{sec:lean-language} by introducing the Lean language and syntax, in order to provide a reader unfamiliar with either theorem provers or Lean with the basic tools needed to understand relevant code snippets.
It will then summarize the scope of the Lean mathematics library \mathlib \cite{themathlibcommunityLeanMathematicalLibrary2020}, and how we intend this work to fit within it.
With these preliminaries in place, it will outline in \cref{sec:existing-formalizations} the approach taken by previous formalizations of Geometric Algebra in other languages, and summarize their strengths and shortcomings.
At this point, we introduce our own Lean formalization, contrasting its foundations to those of prior work, and demonstrating its strengths and weaknesses by example.
We conclude by outlining gaps in \mathlib's API which, if filled, would help to tie together our formalization with other branches of mathematics, as well as how ongoing work in \mathlib will open doors to further formalization goals.

\section{Geometric Algebra}
\label{sec:ga}
Geometric algebra (GA) \footnote{often used interchangeably with Clifford algebra} provides a mechanism to extend a vector space with a metric-dependent product operator, the geometric product.
This product combines and generalizes the properties of the conventional dot and cross product used in 3D euclidean space, while inheriting their geometric interpretations; as an example, for two vectors $a$ and $b$, the product decomposes as
$$
ab = a\cdot b + a \wedge b
$$
where $a\cdot b$ is a scalar and $a \wedge b$ is a bivector analogous to $a \times b$.

A geometric algebra is characterized by its metric, a quadratic form $Q(x)$ assigning a scalar value to the square of each vector $x$.
It is typical\footnote{although not assumed throughout this paper} for our vectors to be a module over $\mathbb{R}$, where the action of the quadratic form can be characterized by its action on a set of orthogonal basis vectors $e_i$.
Specifically, the quadratic form can be encoded a 3-tuple \enquote{signature} of natural numbers $(p,q,r)$, indicating that $p$ of the basis vectors satisfy $Q(e_i) = 1$, $q$ satisfy $Q(e_i) = -1$, and $r$ satisfy $Q(e_i) = 0$.
For these real vector spaces, we use the notation $\mathcal{G}(\mathbb{R}^{p,q,r})$ to denote an algebra of a given signature.
Some commonly used geometric algebras for 3D geometry are $\mathcal{G}(\mathbb{R}^{4, 1, 0})$, Conformal Geometric Algebra (CGA); and $\mathcal{G}(\mathbb{R}^{3, 0, 1})$, Plane based Geometric Algebra (PGA).
These are particularly valuable to applications in computer graphics and robotics, as \enquote{blade} elements of the algebra represent 3D primitives like lines, planes, and circles, while \enquote{rotor} elements of the algebra represent translations, rotations, and reflections.
Many other familiar algebras are isomorphic to a geometric algebra, like the complex numbers $\mathcal{G}(\mathbb{R}^{0,1,0})$ and the exterior algebra $\mathcal{G}(\mathbb{R}^{0,0,n})$.

A key advantage to using GA for geometric problems is that it is rarely necessary to explicitly separate coordinates when doing geometry, as would be done for instance when constructing a rotation matrix or computing the intersection of two planes.
In a numerical software package, internally representing coordinates is unavoidable; but for symbolic manipulation, coordinates can just be a distraction.
For this reason, we would prefer a formalization that avoids coordinates entirely.

\section{Theorem proving software}
Theorem proving software must do more than simply encode and verify mathematical truth; it must make doing so ergonomic for the user, and minimize the inevitable friction when converting a pen and paper argument into a machine-readable one.
The typical workflow for this conversion is split into two parts---translating the theorem statement, and then translating the proof.

The use of dependent type theory (as opposed to the set theory typically used by mathematicians) helps with the theorem statements, as it provides a mechanism to reject nonsensical statements like $1 = (2 < 3)$ (with a type error like \enquote{$2 < 3$ is a Prop, expected an $\mathbb{N}$}).
Flexible notation is also valuable in theorem statements, as mathematics is symbol-heavy; for instance, being able to write $\sum$ instead of \enquote{sum} makes it easier to align paper with screen.

Where the software can really shine though is in the proofs, via interactive and automated theorem proving.
The former is about showing unresolved \enquote{goals} as the proof progresses, to help the user as they write and to guide them what to do next.
The latter is invaluable for discharging goals the user views as trivial, such as $(a + b) + c = (b + c) + a$---while the user could manually apply associativity and commutativity lemmas, automation prevents them having to waste time thinking about this.
The scale of automation can vary from filling in the blanks (converting a proof of $x < x + 1$ into a proof of $2 < 3$), to performing a CAS-like simplification, to applying a sophisticated machine-learning model \cite{polu2020generative}.
Since automation is typically implemented as proof-generation, it doesn't impact the trustworthiness of the prover---the generated proof is subject to the same mechanical scrutiny as a hand-written one.
The line between interactive and automated is frequently blurred.

The Lean 3 language used in this paper is just one of many theorem proving languages, and while it has all of the properties described above, it is not unique in that it does so.
Its use of dependent types is very similar to Coq, and its heavy use of notation similar to Agda.
Its automation tools are currently less powerful than those of Isabelle/HOL, without an equivalent to Isabelle/HOL's powerful \mintinline{isabelle}{sledgehammer} tactic \cite{sledgehammer}.

\section{A quick primer on Lean}
\label{sec:lean-language}

In this section, we will give a very brief introduction to the Lean language, in order to aid with reading the fragments of Lean code\footnote{Note that Lean source code makes heavy use of non-ASCII unicode characters, which some PDF readers are unable to copy to the clipboard faithfully.} in the rest of this paper.
The Lean community website provides a much more in-depth set of reference materials \cite{learning_lean}, as well as an in-browser Lean 3 environment \footnote{Which may in future run a version of Lean incompatible with the code samples in this paper. A compatible version can be ensured by using the configuration in our GitHub repository in \cref{sec:conclusions}.} \cite{lean_web_editor}.

We'll start with a simple definition, showing how to define a value with a given name and type.
Here, we define the name \mintinline{lean}{two} to refer to the natural number (\mintinline{lean}{ℕ}) 2:
\begin{minted}{lean}
-- name : type := value (or "term")
def two : ℕ := 2
\end{minted}
Function \emph{types} are declared with a \mintinline{lean}{→} separating the types of their inputs and outputs.
Function \emph{values} introduce their variable binders using a \mintinline{lean}{λ} followed by variable names.
Here, we define a function that takes a natural number and doubles it:
\begin{minted}{lean}
def double : ℕ → ℕ := λ a, two * a
\end{minted}
For convenience, the language lets us introduce the variable and declare its type in one place; so we could also have written this:
\begin{minted}{lean}
def double (a : ℕ) : ℕ := two * a
\end{minted}
Function application is notated with juxtaposition, which has higher precedence than most operators.
Note that sometimes a \mintinline{lean}{$} is used in place of parentheses.
\begin{minted}{lean}
/- examples of function application -/
def four := double 2
def five := double 2 + 1
def six := double (2 + 1)
def six' := double $ 2 + 1
\end{minted}
What makes Lean a language for theorem proving, and not just a regular functional programming language, is its \mintinline{lean}{Prop} type, which holds mathematical statements.
\begin{minted}{lean}
def likely : Prop := ∃ x : ℕ, x + x = 2
def unlikely : Prop := ∃ x : ℕ, x*x = 2
\end{minted}
Mathematical statements are themselves types (are permitted to appear to the right of a colon), whose values constitute their proofs.
We use the keyword \mintinline{lean}{lemma} instead of \mintinline{lean}{def} when providing proofs, but the syntax is otherwise identical:
\begin{minted}{lean}
lemma likely_proof : likely := ⟨1, rfl⟩
lemma unlikely_proof : unlikely := sorry
\end{minted}
We can prove \mintinline{lean}{likely} by claiming that \mintinline{lean}{1} is a suitable \mintinline{lean}{x}, and proving that \mintinline{lean}{1 + 1 = 2} by definition (\mintinline{lean}{rfl} is a proof that \mintinline{lean}{a = a}).
However, while we are free to make the false statement \mintinline{lean}{unlikely}, Lean does not allow us to prove it.
Instead, it grants us an escape with the \mintinline{lean}{sorry} keyword, which is used to mark proofs as omitted.
Using \mintinline{lean}{sorry} sets a contagious flag that marks a proof and all its dependencies as incomplete, ensuring that the user eventually goes back to fill in the missing proof.
In this paper, we will use \mintinline{lean}{sorry} in code examples to indicate proofs not interesting to the reader, but present in our formalization.

Usually we do not separate our statements and proof as we did for \mintinline{lean}{likely}.
The example below shows a slightly longer proof, using our earlier definition of \mintinline{lean}{double}:
\begin{minted}{lean}
lemma double_is_add_self : ∀ a, double a = a + a :=
  by { intro a, apply two_mul }
\end{minted}
This can be read as a function, that takes a number \mintinline{lean}{a} and emits a proof of a statement about that specific \mintinline{lean}{a}.
This is what it means for Lean to be dependently-typed; the type of the result of a function can depend on its input.
When the inputs of a function are themselves proofs, this concept of proofs as values leads naturally to the Howard-Curry correspondence; an implication of the form $P \implies Q$ is represented as a function that converts proofs of $P$ into proofs of $Q$.

The second line of this proof enters \emph{tactic mode} using the \mintinline{lean}{by} keyword.
Once within this mode, tactics like \mintinline{lean}{intro} and \mintinline{lean}{apply} can be used.
These tactics are often themselves Lean programs, and this mode is what makes Lean an \enquote{interactive} theorem prover.
When in this mode, compatible text editors will show what is left to be proven after each tactic has been applied.

The logical foundation of Lean is the Calculus of Inductive Constructions.
The following example demonstrates a typical inductive construction; an inductive type used to represent the logical or of two propositions:
\begin{minted}{lean}
/-- a pedagogical copy of the built-in `or` -/
inductive my_or (P Q : Prop) : Prop
| left (p : P) : my_or
| right (q : Q) : my_or
\end{minted}
\label{code:my_or}

This reads as \enquote{there are two ways to construct a proof of $P \vee Q$; by providing a proof of $P$ (\mintinline{lean}{my_or.left p}), and by providing a proof of $Q$ (\mintinline{lean}{my_or.right q})}.
From this definition, Lean provides us with an induction principle \mintinline{lean}{my_or.rec} which allows this procedure to be reversed, as \enquote{To prove\footnote{Here and in Lean, $\vee$ is the logical disjunction, not the regressive product of GA} $P \vee Q \to R$, it is sufficient to prove $P \to R$ and $Q \to R$}.

These \enquote{inductive} types are used to build almost every object needed in formalized mathematics and computer science; examples include the integers, lists, logic operators, and existential quantifiers.
Sometimes we need one more fundamental building block, \enquote{quotient} types.
These allow associating an equivalence relation with a type, for instance to ensure that $\frac{1}{2} = \frac{2}{4}$.
Lean supports these too, and we will see more of them in \cref{sec:ours:quotient}.

\section{Lean's Mathematical Library, \mathlib}
\label{sec:lean_mathlib}

By itself, Lean is very much \enquote{batteries not included}.
Its standard library is not opinionated on whether you use it for mathematics or software verification, and as a result comes with little beyond basic data types (e.g. \mintinline{lean}{int}, \mintinline{lean}{list}) and constructive logic.
Notably, it does not contain proofs of statements like \enquote{the integers form a ring}, nor even the definition of \enquote{ring} required to make such a statement!
This is not a short-coming in Lean---it's just a choice of division of responsibility.

The mathematically interesting statements come from \mathlib, which is \enquote{a community-driven effort to build a unified library of mathematics formalized in the Lean proof assistant} \cite{themathlibcommunityLeanMathematicalLibrary2020}, spanning topics like linear algebra, topology, and analysis.
The library is constantly growing \cite{mathlib_stats}, with over 100 contributors whose backgrounds are widely spread  across both fields and seniority, with valuable contributions coming from undergraduates and professors alike.
Perhaps a particularly notable property of \mathlib is that one of its goals is to formalize the entirety of a particular undergraduate mathematics curriculum \cite{mathlib_undergrad_math}.

The coherency and breadth of this library makes formalization work in Lean particularly attractive, as new structures can often be built as a thin layer on top of existing structures.
Furthermore, it reduces the effort needed to bring formalizations in different branches of mathematics together---for instance, in the process of formalizing statements about the wedge product in this work (\cref{sec:ours:wedge}), the \mathlib definition of the matrix determinant was shown to be an alternating map.

Lean is not unique in having a standard library of mathematics---two obvious contenders are the math-comp of Coq, and the stdlib of Agda.
A thorough comparison of the three libraries is beyond the scope of this paper, but a very rough estimate of breadth and accessibility to new users can be obtained by counting the lines of code, comments, and contributors, as shown in \cref{table:lib-comparison}.
This estimate is a poor one, as differences between the languages themselves could easily result in the same idea taking a different number of lines to express---but it's sufficient to demonstrate that Lean is at least on-par with its competitors.
For reference we also include Hol-light and Isabelle's \enquote{Archive of Formal Proofs} (AFP) in \cref{table:lib-comparison}, but as noted in the table caption these are not suited for direct comparison.

\begin{table}
\caption{Comparison of metrics between standard libraries of popular provers using data from their GitHub repositories as of July 2021.}
\begin{threeparttable}
\begin{tabular}{l|cccc}
Language & Library & Source lines\tnote{1} & Comment lines\tnote{1} & Contributors\tnote{2} \\
\hline
Lean & \mathlib & 388k & 100k & 159 \\
Agda & agda-stdlib & 66.9k & 20.4k & 98 \\
Coq & math-comp & 95.9k & 7.6k & 35 \\
\hline
Hol-light\tnote{3} &  & 707k & 40.4k & 8 \\
Isabelle & AFP\tnote{4} & 4.26M & 108k & 386 \\
\end{tabular}
\begin{tablenotes}
\footnotesize
\item[1] Counted using the \texttt{scc} program.
\item[2] According to \texttt{git}. Some projects do not record all authors here.
\item[3] The standard library and language are hard to distinguish, so the metrics include both.
\item[4] The \enquote{Archive of Formal Proofs} is not a standard library but a collection of libraries, and as such is larger but less interlinked.
\end{tablenotes}
\end{threeparttable}
\label{table:lib-comparison}
\end{table}

Our intent in formalizing Geometric Algebra is for our formalization to not only interface well with \mathlib, but also for certain portions to ultimately end up a part of it.
Contributing code into \mathlib ensures its ongoing maintenance \cite{benzmuller_maintaining_2020}, and has numerous advantages: community review by Lean experts, automated detection of bad practices by software tools, and generated documentation published to the web.
Perhaps most important though is that the community as a whole takes on the responsibility of keeping our contributions compatible with the rest of \mathlib as it evolves.
The quality bar for inclusion into \mathlib is high, and as is common in software development, the review process favors lots of small contributions over one giant contribution.
In practice this means it works best to develop larger formalizations outside \mathlib, while in parallel continually shifting parts of their foundations back into \mathlib; this allow rapid iteration unimpeded by reviewer delay, while still insuring against the formalization diverging irrecoverably from \mathlib.

In the rest of this section, we will introduce the elementary parts of \mathlib that will be essential to both translating existing formalizations and constructing our own.

\subsection{Algebraic structures}
Some of the simplest definitions provided by \mathlib are those of the algebraic structure hierarchy---statements like \enquote{the type $M$ is a monoid if it is a semi-group, has an identity $1$, and $\forall m : M, m \cdot 1 = 1 \cdot m = 1$}.
These definition are intended to be used through type-classes, which are a mechanism for associating structure with a type.
This mechanism comes in two parts: a syntax using \mintinline{lean}{[]} for \emph{finding} type-classes,
\begin{minted}{lean}
lemma difference_of_squares {R : Type*} [ring R] (a b : R) :
  (a + b) * (a - b) = a * a - b * b := sorry
\end{minted}
and the \mintinline{lean}{instance} keyword for registering an instance of a type-class to custom types.
\begin{minted}{lean}
instance : ring my_type := sorry
\end{minted}
Combined, these declarations allow our \mintinline{lean}{difference_of_squares} lemma to be applied to values in \mintinline{lean}{mytype}.

Of particular interest to formalization of geometric algebra are the \mintinline{lean}{module R V} and \mintinline{lean}{algebra R A} type-classes\footnote{Which themselves require \emph{at least} \mintinline{lean}{[semiring R] [add_comm_monoid V]} and \mintinline{lean}{[comm_semiring R] [semiring A]}, respectively.}, which respectively describe an $R$-module structure on $V$ (roughly a vector space) and an $R$-algebra structure on $A$.
The former is of interest as it characterizes an arbitrary vector space over which we wish to define the geometric algebra, while the latter is a minimum requirement which our definition must meet in order to be considered useful.

\subsection{Homomorphisms}
\label{sec:mathlib:homs}
In mathematics, it is frequently useful to talk about structure-preserving morphisms, such as an $R$-linear map; a function $f(x) : A \to B$ that $f(x + y) = f(x) + f(y)$ and $\forall c \in R, f(cx) = cf(x)$.
In \mathlib this is represented by the type \mintinline{lean}{linear_map R A B} that \enquote{bundles} the map itself with proofs of the aforementioned properties.
Since such bundled maps are quite common in \mathlib, shorthand notation is provided.
\Cref{table:mathlib-homs} summarizes \mathlib's bundled maps and their notations used in this paper.
\begin{table}
\caption{Notation for homomorphisms in \mathlib}
\begin{tabular}{ll p{0.6\linewidth}}
	Notation & Name & Properties of the map $A \to B$ \\
	\hline
	\mintinline{lean}{A →ₗ[R] B} & \mintinline{lean}{linear_map R A B} & Preserves $+$, $0$, and scaling by \mintinline{lean}{R} \\
	\mintinline{lean}{A →ₐ[R] B} & \mintinline{lean}{alg_hom R A B} & Preserves $+$, $\times$, $1$, $0$, and scaling by \mintinline{lean}{R} \\
	\mintinline{lean}{A ≃ B} & \mintinline{lean}{equiv A B} & Has an associated map $f^{-1} : B \to A$ which is a left- and right- inverse to $f$ \\
	\mintinline{lean}{A ≃ₗ[R] B} & \mintinline{lean}{linear_equiv R A B} & The properties of both \mintinline{lean}{A →ₗ[R] B} and \mintinline{lean}{A ≃ B}
\end{tabular}
\label{table:mathlib-homs}
\end{table}


\section{Existing formalizations of GA}
\label{sec:existing-formalizations}
While there has been no published use of Lean to formalize geometric, Clifford, or Grassman-Cayley algebras, there are already formalizations in other theorem proving languages.
To aid the reader, when describing these other formalizations this section presents code snippets translated into their (roughly) equivalent Lean code.

Of particular interest in these existing formalizations is the underlying definition used to define a multivector, and how this definition can be used to define a product of some kind.
While proofs are obviously important to a formalization, every proof has to start with a theorem statement, the expressivity of which is limited by the definitions at hand.

We present them roughly in order of generality.

\subsection{Fixed-dimension representations}
In \cite{tetsuo_ida_new_2016}, a formalization in Isabelle/HOL specific to $\mathcal{G}(\mathbb{R}^3)$ is presented that enumerates all the $r$-vectors explicitly by assigning a name to each grade:
\begin{minted}[autogobble]{lean}
@[ext]
structure multivector :=
(scalar : ℝ) (vector : ℝ × ℝ × ℝ) (bivec : ℝ × ℝ × ℝ) (trivec : ℝ)
\end{minted}
From here, the operations of an abelian group can be defined componentwise:
\begin{minted}{lean}
instance : has_zero multivector := ⟨
{ scalar := 0, vector := 0, bivec := 0, trivec := 0}⟩
instance : has_add multivector := ⟨λ x y,
{ scalar := x.scalar + y.scalar, vector := x.vector + y.vector,
  bivec := x.bivec + y.bivec, trivec := x.trivec + y.trivec}⟩
-- and for `has_sub`, `has_neg`
\end{minted}
and shown to satisfy its axioms using the \mintinline{lean}{ext} lemma which states \enquote{\mintinline{lean}{multivector}s are equal if their four components are equal}:
\begin{minted}{lean}
instance : add_comm_group multivector :=
{ zero := 0, add := (+), sub := has_sub.sub, neg := has_neg.neg,
  add_zero := λ _, ext _ _ (add_zero _) (add_zero _) (add_zero _) (add_zero _),
  ..sorry /- the 4 other axioms are proved the same way -/ }
\end{minted}
Already we can see that this approach inevitably leads to a lot of repetition, as while it would be easy to generalize over the integer grades $0, 1, 2, 3$, it's challenging to generalize over the names \mintinline{lean}{scalar}, \mintinline{lean}{vector}, \mintinline{lean}{bivec}, and \mintinline{lean}{trivec}.

The problem only becomes worse when defining the geometric product, as now we have 4 terms in each of the components. As in \cite{tetsuo_ida_new_2016}, the full definition is omitted below.
\begin{minted}{lean}
instance : has_one multivector := ⟨
{ scalar := 1, vector := 0, bivec := 0, trivec := 0}⟩
instance : has_mul multivector := ⟨λ x y,
{ scalar := x.scalar * y.scalar + dot_product x.vector y.vector
            - dot_product x.bivec y.bivec - x.trivec * y.trivec,
  vector := sorry, bivec := sorry, trivec := sorry}
\end{minted}
From here, \cite{tetsuo_ida_new_2016} goes on to prove that multivectors form a ring, by showing that multiplication associates and distributes, and operates as expected with one.
Again, this would have to be done component-wise, and the trivial but presumably verbose proofs are omitted from \cite{tetsuo_ida_new_2016}.

The tedium of the component-wise definitions and proofs can be reduced by generation from CAS implementations and automation features within the theorem provers, but this scales poorly to more complex statements, which may need to be tackled by hand.
Needless to say, this formalization does not scale at all to other dimensions and signature of algebra, as no part of it is generalized.

\subsection{Recursive tree representations}
\label{sec:theirs:thery}
A convenient way to escape this death-by-cases is to use a recursive definition of a multivector.
This is the approach taken by the Coq formalization of Grassmann-Cayley algebra in \cite{fuchs_formalization_2010} (and resembles the computational approach used by Garamon, \cite{breuils_garamon_2019}).
There, the definition is built as a balanced binary tree, where each branch indicates the presence or absence of a basis blade, and the leafs contain the corresponding coefficient.
\begin{minted}[autogobble]{lean}
def Gₙ : ℕ → Type
| 0       := ℝ           -- a scalar coefficient
| (n + 1) := Gₙ n × Gₙ n  -- the parts without and with $eₙ$
\end{minted}
In this Lean translation, the \mintinline{lean}{|} syntax is used to pattern match against an integer representing the remaining depth of the tree, while \mintinline{lean}{×} captures the branching.
For instance, $a + a_1e_1 + a_2e_2 + a_{12}e_{12}$ is represented as a term of type \mintinline{lean}{Gₙ 2} as $((a, a_2), (a_1, a_{12}))$.

The recursive data definition leads naturally to a recursive operator definition, which resembles the following\footnote{The \mintinline{lean}{̅y₁ᵈ} notation is introduced in \cite{fuchs_formalization_2010}, but not essential to get a feel for how the recursive definitions look}:
\begin{minted}[autogobble]{lean}
reserve infix ` ⋏ `:70
def wedge : Π {n}, Gₙ n → Gₙ n → Gₙ n
| 0       x       y       := (x * y : ℝ)
| (n + 1) ⟨x₁, x₂⟩ ⟨y₁, y₂⟩ := let infix ` ⋏ ` := wedge in
      (x₂ ⋏ y₂, x₁ ⋏ y₂ + x₂ ⋏ ̅y₁ᵈ)
infix ` ⋏ ` := wedge
\end{minted}
Here, the pattern-matching is against the depth of the tree, and for non-root elements, the two branches.
This elegantly avoids having to deal with coefficient-wise proofs, and results in a definition that works on algebras of arbitrary dimension \mintinline{lean}{n}.

While \cite{fuchs_formalization_2010} does not extend to defining a metric or the geometric product it infers, their follow-up work \cite{fuchs:hal-01095495} does so.
This design still shares a shortcoming of the previous design---it imbues the definition with a canonical and orthogonal basis, which is at odds with our goal of being coordinate-free.

\subsection{Indexed coordinate representations}
The HOL light formalization in \cite{li_formalization_2019} offers the same generality as the one in \cref{sec:theirs:thery}, but slightly more abstractly describing a multivector as a set of coefficients indexed by the IDs of its basis vectors.
\begin{minted}[autogobble]{lean}
variables (n : ℕ)

-- mapping from subsets of 1:n to coefficients
abbreviation idx := set (fin n)
def multivector : Type := idx n → ℝ
\end{minted}
\cite{li_formalization_2019} goes on to define the geometric product and various derived products for arbitrary metrics, which means their formalizations can be used for both CGA and PGA.
A rough translation of their generalized product formalization is as follows.
\begin{minted}[autogobble]{lean}
def generic_prod (a b : multivector n) (sgn : idx n → idx n → ℝ) : multivector n :=
∑ ai bi in finset.univ,
  pi.single (ai Δ bi) ((a ai * b bi) * sgn ai bi)
\end{minted}
Here \mintinline{lean}{ai Δ bi} is the symmetric difference.
This is then used to derive the wedge and other products as
\begin{minted}[autogobble]{lean}
def wedge (a b : multivector n) : multivector n :=
prod a b $ λ ai bi,
  if ai ∩ bi ≠ ∅ then 0 else  -- matching blades
  (-1) ^ (finset.card $ (ai.product bi).filter $ λ abj, abj.1 > abj.2)
\end{minted}

Overall, the formalization that follows in \cite{li_formalization_2019} is expansive, covering topics ranging from the existence of inverses to outermorphisms.
However, the initial definition of \mintinline{lean}{multivector} ingrains a preferred choice of orthogonal (finite) basis, which while in line with many numerical packages for Clifford algebras, is at odds with how vector spaces are formalized in \mathlib.
\mathlib's approach is typically axiomatic, introducing explicit sets of basis vectors only when needed; often, only the proof of the existence of a set of basis vectors is used.

To mesh well with \mathlib, our formalization will need to support algebras over a variety of vector spaces, not just those with coefficients in $\mathbb{R}^n$.
We are of course free to take on extra assumptions should we need them (such as the scalars forming a field, or the dimension of the space being finite), but by making the initial definitions more general we leave the door open to future researchers interested in other algebras which do not satisfy these assumptions.

\section{Our formalization of GA}
\label{sec:ours}
In this section, we present a variety of definitions and theorems for GA that we formalized in Lean.
For reasons described in \cref{sec:lean_mathlib}, some of these formalizations have been integrated into \mathlib, while the rest are available in our source code repository (linked in \cref{sec:conclusions}).
We won't attempt to distinguish where the line is drawn, as ultimately the result is the same; users of our source repository will automatically get a compatible version of \mathlib.

\subsection{Remarks on type theory}
\label{section:algebra-map-iota}
While it is typical on paper to avoid distinguishing \enquote{the multivectors of grade zero} from \enquote{the scalars}, or \enquote{the multivectors of grade 1} from \enquote{the vectors}, the strict dependent typing of Lean forces us to treat these separately.
Instead of saying that the multivectors $\mathcal{G}(V)$ \enquote{contain} the scalars $R$ and vectors $V$, we need to provide explicit mappings between these distinct types.
Respectively, these come in the form of a ring homomorphism $\mathtt{algebra\_map} : R \to \mathcal{G}(V)$ and a linear map $\mathtt{\iota} : V \to \mathcal{G}(V)$\footnote{The names of which are taken from \mathlib conventions}.

Typically we would assume these mappings are injective, but we can obtain many results without needing to do so.
As it turns out, the definitions outlined below permit these maps to be non-injective, for particularly pernicious choices of $R$ and $V$\cite{mathoverflow-87958}\footnote{A partial formalization of this result was shown in Lean.}.

\subsection{The quotient definition}
\label{sec:ours:quotient}
A Clifford algebra over the vector space $V$ with quadratic form $Q : V \to R$ can be defined as a quotient of the tensor algebra $T(V)$ by the two-sided ideal $I_Q$ generated from the set $\{ v \otimes v - Q(v) \:\vert\: v \in V \}$.
Generating an ideal from this set amounts to taking the smallest superset of it that is closed under left- and right- multiplication and addition by elements of $T(V)$.
When we take a quotient by an ideal, we are saying that two elements are considered equivalent if their subtraction is in that ideal.
It follows then that we have $v \otimes v \approx Q(v)$, and therefore our construction ensures that vectors square to scalars.

As of writing, \mathlib does not have direct support for two-sided ideals, but it does support the equivalent operation of taking the quotient by a suitable closure of a relation like $v \otimes v \approx Q(v)$.
As such, the quotient definition still translates fairly naturally into Lean:
\begin{minted}[autogobble]{lean}
variables {R : Type*} [comm_ring R]
variables {V : Type*} [add_comm_group V] [module R V]
variables (Q : quadratic_form R V)

inductive rel : tensor_algebra R V → tensor_algebra R V → Prop
| of (v : V) : rel
    (tensor_algebra.ι R v * tensor_algebra.ι R v)
    (algebra_map R (tensor_algebra R V) (Q v))

/-- `clifford_algebra Q' is the algebra over `V` with metric `Q` -/
@[derive [ring, algebra R]]
def clifford_algebra := ring_quot (clifford_algebra.rel Q)

/-- `ι Q v' is the embedding of the vector `v : V' into the Clifford
algebra with quadratic form `Q`. -/
def clifford_algebra.ι : V →ₗ[R] clifford_algebra Q :=
(ring_quot.mk_alg_hom R _).to_linear_map.comp (tensor_algebra.ι R)
\end{minted}
What makes this definition particularly attractive is that thanks to the \mintinline{lean}{@[derive ...]} attribute (a Lean metaprogram built into \mathlib), the \mintinline{lean}{ring} and \mintinline{lean}{algebra} structure can be automatically proved by Lean using its knowledge of \mintinline{lean}{ring_quot}.
Note that it is this \mintinline{lean}{ring (clifford_algebra Q)} structure which provides the geometric product \mintinline{lean}{*}!

The operations described in \cref{section:algebra-map-iota} also fall out with minimal effort:
the derived \mintinline{lean}{algebra R (clifford_algebra Q)} structure provides the map from the scalars \mintinline{lean}{algebra_map R _ r}; while
the map from the vectors \mintinline{lean}{clifford_algebra.ι Q v} is obtained by first mapping the vectors into the tensor algebra (using \mintinline{lean}{tensor_algebra.ι R}, the analogous $\iota$ for the tensor algebra), and then embedding them within the quotient using the API around \mintinline{lean}{ring_quot}.
Putting these together, we can verify that our construction does indeed square vectors to scalars:
\begin{minted}[autogobble]{lean}
theorem ι_sq_scalar (v : V) : ι Q v * ι Q v = algebra_map R _ (Q v) :=
begin
  erw [←alg_hom.map_mul, ring_quot.mk_alg_hom_rel R (rel.of v), alg_hom.commutes],
  refl,
end
\end{minted}

The remaining code to explain is our \mintinline{lean}{rel} definition, which demonstrates how \mintinline{lean}{inductive} types can be use for propositions, as was introduced briefly in \cref{code:my_or}.
Here, we declare \mintinline{lean}{rel T₁ T₂} as a proposition over pairs of elements in $T(V)$, but provide only one way to construct it, \mintinline{lean}{rel.of v}.
In essence, this means that a proof of \mintinline{lean}{rel T₁ T₂} is a proof that there exists some $v$ such that $T_1 = \operatorname{\iota}(v) \otimes \operatorname{\iota}(v)$ and $T_2 = Q(v)$.
Note that \mathlib uses \mintinline{lean}{*} not \mintinline{lean}{⊗} for the product in the tensor algebra.

While not written by the authors, a better understanding of the power of inductive types can be obtained by looking inside \mathlib's definition of \mintinline{lean}{ring_quot}.
The inductive type used by this definition to extend our \mintinline{lean}{rel} definition to its closure under ring operations is roughly as follows:
\begin{minted}[autogobble]{lean}
inductive crel (rel : R → R → Prop) : R → R → Prop
| of ⦃x y : R⦄ (h : rel x y) : crel x y
| add_left ⦃a b c⦄ : crel a b → crel (a + c) (b + c)
| mul_left ⦃a b c⦄ : crel a b → crel (a * c) (b * c)
| mul_right ⦃a b c⦄ : crel b c → crel (a * b) (a * c)
\end{minted}
This reads as \enquote{A pair of elements are satisfied by the closure of a relation if:}
\begin{description}
    \item[\mintinline{lean}{of}]\enquote{They satisfy the relation \mintinline{lean}{rel}}
    \item[\mintinline{lean}{add_left}]\enquote{They can each be split into an addition with the same right operand \mintinline{lean}{c}, and with left operands satisfying the closure of the relation.}
    \item[\mintinline{lean}{mul_left}, \mintinline{lean}{mul_right}] \enquote{They can each be split into an multiplication with the same left/right operand, and with a right/left operand satisfying the closure of the relation.}
\end{description}
One might remark that something still seems odd about this relation, as neither this \mintinline{lean}{crel} nor our \mintinline{lean}{rel} imply reflexivity, symmetry, or transitivity --- however, these properties follow from the axiomatization of Lean's \mintinline{lean}{quot} type, and are provided as part of the Lean prelude as \mintinline{lean}{quot.exact : quot.mk r a = quot.mk r b → eqv_gen r a b}.

\subsection{The universal property}
\label{section:algebra-map-universal-property}
While convenient to define, the quotient can be hard to state further definitions and prove theorems about.
When defining operations over a quotient, the approach is almost always to operate on the data within the quotient, and then prove that for any operands that are considered equal under the quotient, the output of the operation is unchanged.

Such proofs can be very challenging, especially given some short-comings in Lean when it comes to recursing over nested inductive types (such as \mintinline{lean}{crel} which wraps our \mintinline{lean}{rel} above).
Another approach is to work with the universal property from category theory.
The universal property of the Clifford algebra is shown diagrammatically in \cref{eq:univ}---given a linear map $f$ from $V$ to an algebra $A$ satisfying $f(v)f(v) = Q(v)\cdot 1_{A}$, there is a unique addition-and-multiplication-preserving map from $\mathcal{G}(Q)$ to $A$ that satisfies $f = (\operatorname{lift} f) \circ \iota $.
Since $\operatorname{lift}$ provides a unique mapping, it has an inverse and we can also state this property in reverse in terms of $g$.
\begin{equation}
\begin{tikzcd}
\mathcal{G}(Q) \arrow[r, "\operatorname{lift} f = g"] & A \\
V \arrow[ru, "f = \operatorname{lift}^{-1} g"'] \arrow[u, "\iota" description]
\end{tikzcd}\label{eq:univ}
\end{equation}

The universal property can be stated in Lean as
\begin{minted}{lean}
def lift :
  {f : V →ₗ[R] A // ∀ v, f v * f v = algebra_map _ _ (Q v)}
    ≃ (clifford_algebra Q →ₐ[R] A) := sorry
\end{minted}
which with the help of \cref{table:mathlib-homs}, reads piecewise as:
\begin{description}
    \item[\mintinline{lean}{lift : _ ≃ _}] \enquote{lift} is an equivalence between...
    \item[\mintinline{lean}{V →ₗ[R] A}]... $R$-linear maps from the vector space $V$ to the algebra $A$...
    \item[\mintinline{lean}{{f : _ // ∀ v, f v * f v = algebra_map _ _ (Q v)}}] ... whose output squares to the metric ...
    \item[\mintinline{lean}{clifford_algebra Q →ₐ[R] A}] ... and maps between $\mathcal{G}(Q)$ and $A$ which are $R$-linear and preserve multiplication.
\end{description}
The implementation of \mintinline{lean}{lift} (as opposed to just its type) is not included in this paper, but consists largely of invoking uninteresting machinery around \mintinline{lean}{ring_quot} already in \mathlib.
This machinery within \mathlib encapsulates the task of working under the quotient, so that we don't have to

Note that if we apply this equivalence in reverse to the identity algebra automorphism $\mathcal{G}(Q) \to \mathcal{G}(Q)$ (that is, evaluate \mintinline{lean}{(lift Q).symm (alg_hom.id R _)}), then we recover a linear map from $V$ to $\mathcal{G}(Q)$ whose results square to scalars. Intuitively\footnote{But thankfully, also proven by Lean!}, this is $\iota(v)$.

The type of \mintinline{lean}{lift} alone is enough for us to prove statements like $(\operatorname{lift} f) (a + b * c) = f(a) + f(b)f(c)$, while its composition with $\iota$ gives us the remaining interesting properties.

For the rest of our formalization, our proofs and definitions depend only on the properties of \mintinline{lean}{lift} and \mintinline{lean}{ι}, and never on the quotient construction from \cref{sec:ours:quotient}.
In principle, this would enable future work to transfer proofs of theorems about our construction to any other construction, such as those in \cref{sec:existing-formalizations}, provided that those definitions are equipped with their own \mintinline{lean}{lift} and \mintinline{lean}{ι}.

\subsection{Conjugations}

Beyond the identity mapping \mintinline{lean}{(lift Q) (ι Q) = alg_hom.id R _}, the next simplest operation we can use it to define is the grade involution $\hat X$, which flips the sign of every component vector. In Lean, we write that
\begin{minted}{lean}
def involute : clifford_algebra Q →ₐ[R] clifford_algebra Q :=
clifford_algebra.lift Q ⟨-(ι Q), λ v, by simp⟩
\end{minted}
where \mintinline{lean}{by simp} is producing the trivial proof that \mintinline{lean}{-(ι Q v)*-(ι Q v) = (ι Q v)*(ι Q v)}.
Just as we saw earlier with \mintinline{lean}{lift}, the type of \mintinline{lean}{involute} alone is enough to prove statements about involutions of addition and multiplication.
This time, we also get from the type the fact that involute acts as the identity on scalars.
All that remains is to prove that involution negates vectors, which get almost for free using some \mintinline{lean}{@[simp]} lemmas about the universal property (not shown):
\begin{minted}{lean}
@[simp] lemma involute_ι (v : V) : involute (ι Q v) = -ι Q v :=
by simp [involute]
\end{minted}
To check that the Lean simplifier is suitably trained for \mintinline{lean}{involute}, we show that taking the involution of a product of $n$ vectors (\mintinline{lean}{(l.map $ ι Q).prod}) is equivalent to scaling by $(-1)^{n}$:
\begin{minted}{lean}
lemma involute_prod_map_ι : ∀ l : list M,
  involute (l.map $ ι Q).prod = ((-1 : R)^l.length) • (l.map $ ι Q).prod
| [] := by simp
| (x :: xs) := by simp [pow_add, involute_prod_map_ι xs]
\end{minted}

A similar approach can be performed to define grade reversion $\tilde X$, although this time instead of inserting a minus sign for each vector, we need to flip the multiplication order.
\mathlib provides us an \mintinline{lean}{op : X → Xᵒᵖ} mapping for exactly that, which by definition satisfies \mintinline{lean}{op x * op y = op (y * x)}. Applying this to each of our vectors will then give \mintinline{lean}{op(reverse(x))}, which we can map back into \mintinline{lean}{X} with \mintinline{lean}{unop : Xᵒᵖ → X}.
It can be trivially shown that this mapping is linear and invertible (\mintinline{lean}{opposite.op_linear_equiv : X ≃ₗ[R] Xᵒᵖ}), which combined with some rather ugly boilerplate gives us a complete definition for grade reversal.
\begin{minted}{lean}
def reverse : clifford_algebra Q →ₗ[R] clifford_algebra Q :=
(op_linear_equiv R).symm.to_linear_map.comp (
  clifford_algebra.lift Q ⟨(opposite.op_linear_equiv R).to_linear_map.comp (ι Q),
    λ m, unop_injective $ by simp⟩).to_linear_map
\end{minted}
Unlike \mintinline{lean}{involute}, \mintinline{lean}{reverse} is only a  \mintinline{lean}{→ₗ[R]} and not a \mintinline{lean}{→ₐ[R]}, so we need to prove how it acts on scalars and multiplication ourselves.
Once again, \mintinline{lean}{simp} makes short work of this.
\begin{minted}{lean}
@[simp] lemma reverse.map_mul (a b : clifford_algebra Q) :
  reverse (a * b) = reverse b * reverse a :=
by simp [reverse]

@[simp] lemma reverse.commutes (r : R) :
  reverse (algebra_map R (clifford_algebra Q) r) = algebra_map R _ r :=
by simp [reverse]
\end{minted}

While the above definitions of \mintinline{lean}{involute} and \mintinline{lean}{reverse} give us proofs about their operations on sums and products essentially for free, they miss one key property of these conjugations; the fact that they are involutive, \mintinline{lean}{reverse (reverse x) = x}.
Proving these becomes much easier once we build an inductive principle.

\subsection{Induction}
\label{sec:ours:induction}
The induction principle we seek can be stated \enquote{If a property \mintinline{lean}{P} holds for the \mintinline{lean}{algebra_map} of \mintinline{lean}{r : R} and the \mintinline{lean}{ι} of \mintinline{lean}{v : V} into \mintinline{lean}{clifford_algebra Q}, and is
preserved under addition and multiplication, then it holds for all of \mintinline{lean}{clifford_algebra Q}}.
Perhaps surprisingly, the universal property alone is enough for us to construct this principle.
An a outline of the approach is:
\begin{enumerate}
    \item Show that collectively, the inputs to our inductive principle define a \mintinline{lean}{subalgebra} (\mintinline{lean}{s}) of precisely the elements that satisfy \mintinline{lean}{P}; that is, a subset of the full Clifford algebra which contains zero and one and is closed under addition and multiplication. By doing this, Lean provides us with a type \mintinline{lean}{↥s} which bundles each element of the subalgebra with a proof that it belongs to that subalgebra.
    \item Restrict the codomain of \mintinline{lean}{ι Q} to \mintinline{lean}{↥s}, and show that doing so still preserves the fact that vectors square to scalars.
    \item Lift this restricted map from \mintinline{lean}{V → ↥s} to \mintinline{lean}{clifford_algebra Q → ↥s}.
    \item Apply this lifted map to our input \mintinline{lean}{a}. We show that the value part of this lifting is just \mintinline{lean}{a}, meaning the proof part is \mintinline{lean}{C a}, our proof for an arbitrary element.
\end{enumerate}
The full Lean implementation is shown below.
\begin{minted}{lean}
@[elab_as_eliminator]
lemma induction {P : clifford_algebra Q → Prop}
  (h_grade0 : ∀ r, P (algebra_map R (clifford_algebra Q) r))
  (h_grade1 : ∀ x, P (ι Q x))
  (h_mul : ∀ a b, P a → P b → P (a * b))
  (h_add : ∀ a b, P a → P b → P (a + b))
  (a : clifford_algebra Q) :
  P a :=
begin
  -- the arguments are enough to construct a subalgebra, and a mapping into it from M
  let s : subalgebra R (clifford_algebra Q) :=
  { carrier := P, mul_mem' := h_mul, add_mem' := h_add, algebra_map_mem' := h_grade0 },
  let of : clifford_hom Q ↥s :=
  ⟨ (ι Q).cod_restrict s.to_submodule h_grade1,
    λ m, subtype.eq $ ι_sq_scalar Q m ⟩,
  -- the mapping through the subalgebra is the identity
  have of_id : alg_hom.id R (clifford_algebra Q) = s.val.comp (lift Q of),
  { ext,
    simp [of], },
  -- finding a proof is finding an element of the subalgebra
  convert subtype.prop (lift Q of a),
  exact alg_hom.congr_fun of_id a,
end
\end{minted}
Armed with our new hammer, we find a lot of nail-like lemmas we can easily prove.
Below, we show that involution and reverse commute, and each is involutive; that is, $\tilde{\hat x} = \hat{\tilde x}$, $\hat{\hat x} = x$, and $\tilde {\tilde x} = x$.
\begin{minted}{lean}
@[simp] lemma reverse_involute_commute :
  function.commute (reverse : _ → clifford_algebra Q) involute :=
λ x, by induction x using clifford_algebra.induction; simp [*]

@[simp] lemma involute_involutive :
  function.involutive (involute : _ → clifford_algebra Q) :=
λ x, by induction x using clifford_algebra.induction; simp [*]

@[simp] lemma reverse_involutive :
  function.involutive (reverse : _ → clifford_algebra Q) :=
λ x, by induction x using clifford_algebra.induction; simp [*]
\end{minted}

\subsection{\texorpdfstring{$\mathbb{Z}_2$}{ℤ₂}-grading}
\label{sec:ours-z2-grading}
An algebra $A$ is said to be graded by an additive monoid $I$ if there exists a family of $I$-indexed submodules $A_i$ such that $x \in A_i,\ y \in A_j \to xy \in A_{i + j}$, $1 \in A_0$, and every element $a \in A$ has a unique decomposition\footnote{Which corresponds to requiring that the submodules $A_i$ span the space, and they satisfy some notion of disjointness.} $a = \sum_i a_i$ where $a_i \in A_i$.
If we adjust this definition to use a wedge product instead of a regular product, then we can obtain the familiar $\mathbb{N}$-grading used in Geometric Algebra.
However, we haven't formalized the wedge product yet!

Instead, we can show that the Clifford algebra permits a $\mathbb{Z}_2$ grading, that is a grading by the integers modulo two.
Intuitively, this maps to the notion of even- and odd-graded $r$-vectors.
Again, we get there via the universal property; this time, mapping into \mintinline{lean}{add_monoid_algebra (clifford_algebra Q) (zmod 2)} where \mintinline{lean}{add_monoid_algebra} is built into \mathlib.
This type can be thought of as somewhat analogous to a polynomial, where the coefficients are the even and odd multivectors themselves, and the variable is either $x = x^3 = \cdots$ (odd) or $1 = x^2 = \cdots$ (even), where multiplication behaves as addition does in $\mathbb{Z}_2$.
We write this in Lean as follows, where the name \mintinline{lean}{single} indicates constructing an object of a single grade.
\begin{minted}{lean}
def grades' :
  (clifford_algebra Q) →ₐ[R] add_monoid_algebra (clifford_algebra Q) (zmod 2) :=
(lift Q : _) ⟨
  -- vectors are grade 1
  (add_monoid_algebra.lsingle 1).comp (ι Q),
  -- this requires 1 + 1 = 0, which is why we use `zmod 2` as our grading
  λ x, by { simp [add_monoid_algebra.single_mul_single], congr }⟩
\end{minted}
Thanks to the universal property, and the fact it produces an \mintinline{lean}{→ₐ[R]} for us, we get obvious properties like \mintinline{lean}{grades' (r + s) i = grades' r i + grades' s i} for free.

To finish this definition of the grading, we show that there is \mintinline{lean}{→ₐ[R]} in the other direction forming a left inverse (by simply summing components), and we show that the grades are disjoint.
We state this latter fact as
\begin{minted}{lean}
@[simp]
lemma grades'.map_grades' (x : clifford_algebra Q) (i : zmod 2) :
  let xi := (grades' x i) in grades' xi = finsupp.single i xi := sorry
\end{minted}
which reads \enquote{If you take the grade $i$ part of $x$ ($x_i$), and then split $x_i$ into its constituent grades, it will consistent of a single grade $i$}.
The Lean proof is omitted in this paper for brevity, but it once again falls out after application of the induction principle from \cref{sec:ours:induction}.

\subsection{Versors}
The versors can be described as a set of multivectors closed under multiplication (\mintinline{lean}{submonoid}) and scaling, generated from the set of vectors (\mintinline{lean}{set.range (ι Q)}). We write this formally as:
\begin{minted}{lean}
def versors := center_submonoid.closure (set.range (ι Q))
\end{minted}
where \mintinline{lean}{center_submonoid} is a wrapper for a set of elements, built on top of the \mintinline{lean}{submonoid} and \mintinline{lean}{sub_mul_action} structures in \mathlib that respectively carry proofs of closure under multiplication and scaling.

Lean provides some useful syntax for working with sub-objects like this.
Instead of working with \mintinline{lean}{(x : clifford_algebra Q) (hx : x ∈ versors Q)}, we can write as a shorthand \mintinline{lean}{(v : ↥versors Q)}.
The advantage of \enquote{bundling} the multivector with its proof of being a versor like this is that operations \mintinline{lean}{v * w} and \mintinline{lean}{k • v} can be defined to automatically produce a bundle containing a proof that their result is also a versor; and indeed, \mintinline{lean}{center_submonoid} does just that.

Armed with our definition, the next step is to once again construct an induction principle. Most of the heavy lifting is done by the \mintinline{lean}{submonoid.closure_induction'} principle in \mathlib, which was developed as part of this paper. The statement of this principle is:

\begin{minted}{lean}
/-- If a statement `C` is true for all scalars, all vectors, and all products
of versors which each satisfy `C`, then it is true for all versors. -/
lemma induction_on {C : versors Q → Prop} (v : versors Q)
  (h_scalars : ∀ r : R, C ⟨algebra_map _ _ r, (versors Q).algebra_map_mem r⟩)
  (h_vectors : ∀ m, C ⟨ι Q m, ι_mem m⟩)
  (h_mul : ∀ a b, C a → C b → C (a * b)) :
  C v := sorry
\end{minted}

The induction principle lets us once again bash out some useful statements with uninteresting proofs.
This time, we scratch the surface of Lean's meta-programming framework to avoid repeating a trivial proof:
\begin{minted}{lean}
/-- A simple macro tactic that we can reuse between proofs -/
meta def inv_rev_tac : tactic unit :=
`[apply induction_on v,
  { intro r, simp, },
  { intro m, simp, },
  { intros a b ha hb, simp [(versors Q).mul_mem, ha, hb] }]

/-- Involute of a versor is a versor -/
@[simp] lemma involute_mem (v : versors Q) :
  involute (v : clifford_algebra Q) ∈ versors Q := by inv_rev_tac

/-- Reverse of a versor is a versor -/
@[simp] lemma reverse_mem (v : versors Q) :
  reverse (v : clifford_algebra Q) ∈ versors Q := by inv_rev_tac
\end{minted}

A more interesting application of our induction principle is to prove that the product of a versor and its reverse is a scalar:
\begin{minted}{lean}
/-- A versor times its reverse is a scalar -/
lemma mul_self_reverse (v : versors Q) :
  ∃ r : R, (v : clifford_algebra Q) * reverse (v : clifford_algebra Q) = ↑ₐr :=
begin
  with_cases { apply induction_on v },
  case h_grade0 : r {
    refine ⟨r * r, _⟩,
    simp },
  case h_grade1 : m {
    refine ⟨Q m, _⟩,
    simp },
  case h_mul : x y {
    rintros ⟨(qx : R), hx⟩ (⟨qy : R), hy⟩, -- results for `x` and `y` by induction
    refine ⟨qx * qy, _⟩,
    simp only [reverse_mul, submonoid.coe_mul, ring_hom.map_mul],
    rw [mul_assoc ↑x, ←mul_assoc ↑y, hy, algebra.commutes, ←mul_assoc, hx], }
end
\end{minted}
Here we use some more verbose Lean syntax to clearly indicate each branch of the induction.
From here, we go on to show that versors have an inverse, and that for a non-trivial algebra over a field in an anistropic metric, they form a group with zero (i.e. all elements but zero have an inverse):
\begin{minted}{lean}
instance
  {K} [field K] {V} [add_comm_group V] [module K V]
  {Q : quadratic_form K V} [nontrivial (clifford_algebra Q)] [f : fact Q.anisotropic] :
    group_with_zero (versors Q) := sorry
\end{minted}

\subsection{Wedge products and \texorpdfstring{$\mathbb{N}$}{ℕ}-grading}
\label{sec:ours:wedge}
Formalizing the wedge product of $n$ vectors can be done as follows,
\begin{minted}{lean}
/-- A wedge product of n vectors. Note this does not define the wedge product of arbitrary multivectors. -/
def ι_wedge (n : ℕ) [invertible (n.factorial : R)] :
  alternating_map R M (clifford_algebra Q) (fin n) :=
⅟(n! : R) •
  ((mk_pi_algebra_fin R n _).comp_linear_map (λ i, ι Q)).alternatization
\end{minted}
The definitions and API for alternating maps, and crucially their construction from the alternatization of a multilinear map (\mintinline{lean}{multilinear_map.alternatization}) were some of the many contributions to \mathlib as part of the work in this paper.
Through this API, we can easily prove statements like \mintinline{lean}{∀ a b, ι_wedge _ ![a, b, a] = 0} (i.e. $a \wedge b \wedge a = 0$ for all vectors $a, b$), or more generally that the wedge product of linearly dependent vectors is zero.

Of course, a crucial use of the wedge product is between arbitrary multivectors, not just basis vectors.
The ideal way to define this wedge product would be to find a linear isomorphism \mintinline{lean}{to_ext : clifford_algebra Q ≃ₗ[R] exterior_algebra R V}, and implement \mintinline{lean}{wedge} by transferring the arguments along the isomorphism, computing the product in the exterior algebra, and transferring back:
\begin{minted}{lean}
def wedge (a b : clifford_algebra Q) : clifford_algebra Q :=
  to_ext.symm (a.to_ext * b.to_ext)
\end{minted}
The advantage of this approach is that theorems about multiplication in exterior algebras can be transferred easily to theorems about the wedge product in Clifford algebras.

Our work does not go as far as defining the \mintinline{lean}{to_ext}, although in collaboration with members of the Lean community the authors contributed a definition of \mintinline{lean}{exterior_algebra R V} to \mathlib using a similar approach to that described in \cref{sec:ours:quotient,section:algebra-map-universal-property}.
Constructing the isomorphism \mintinline{lean}{to_ext} from the universal property alone is difficult.

One way to do so would be to first show that the Clifford algebra is a filtered algebra, in that it can be split into subsets $X_0 \subseteq X_1 \subseteq X_2 \subseteq \cdots$ such that $a \in X_i, b \in X_j \implies ab \in X_{i+j}$. From here, the associated graded algebra can be constructed with grades $Y_0 = X_0, Y_{n + 1} = \tfrac{X_{n+1}}{X_n}$ where division indicates the quotient submodule. Finally, a variant of the Poincaré–Birkhoff–Witt theorem can be applied to find the isomorphism.

The definition of a filtered algebra, along with a proof that a Clifford algebra satisfies it with $X_i$ as \enquote{the multivectors of at most grade $i$} was formalized in Lean by the authors.
The associated graded algebra proved challenging to encode in dependent type theory, as the type of each grade becomes a function of the type of the grade before it.
This obstacle is not insurmountable, but the authors feel it best left to future work.

Interestingly, it can be shown that such an isomorphism does not always exist; as mentioned in \cref{section:algebra-map-iota}, there exist Clifford algebras under our definition where the \enquote{algebra map} is not injective.
However, we show in our formalization that the \enquote{algebra map} of the exterior algebra is always injective.
Since only one of the two maps is always injective, these pathological Clifford algebras cannot be isomorphic to the corresponding exterior algebra.

\subsection{Constructing specific algebras}
In \cref{sec:ga}, we draw attention to some particularly useful geometric algebras over the real numbers, which until this point our formalization has been too general to mention.
In this section, we show how to use our general formalization with specific algebras.
It is worth noting that our formalization does not permit us to make statements like \enquote{the complex numbers are a geometric algebra}, instead requiring us to rephrase to \enquote{the complex numbers are isomorphic to a geometric algebra}.
We state this isomorphism in Lean as follows:
\begin{minted}{lean}
/-- The quadratic form sending elements to the negation of their square. -/
noncomputable def complex.Q : quadratic_form ℝ ℝ :=
-quadratic_form.lin_mul_lin linear_map.id linear_map.id

@[simp] lemma complex.Q_apply (r : ℝ) : complex.Q r = -(r*r) := rfl

/-- The clifford algebra with quadratic form `complex.Q` is isomorphic to `ℂ`. -/
def equiv_complex : clifford_algebra complex.Q ≃ₐ[ℝ] ℂ := sorry

lemma equiv_complex_ι : complex (ι complex.Q 1) = complex.I := sorry
\end{minted}
The full details are provided in our source repository, as well as a similar construction for the quaternions.
This choice to work with isomorphisms is pragmatic one; it's consistent with how \mathlib handles the tensor algebra, and it doesn't force us to declare one particular isomorphism as canonical; in the example above, we could just as easily mapped \mintinline{lean}{ι complex.Q 1} to \mintinline{lean}{-complex.I} instead of \mintinline{lean}{complex.I}.

For a more illustrative example than the complex numbers, we will use the rest of this section to outline how to set up a definition of Conformal Geometric Algebra, $\mathcal{G}(\mathbb{R}^{n+1, 1, 0})$, which augments the initial vector space $\mathbb{R}^{n}$ with two extra dimensions spanned by the orthogonal vectors $e_{+}$ and $e_{-}$ where $e_{+}^2 = 1$ and $e_{-}^2 = -1$.
In practice it can be algebraically convenient to span these extra dimensions instead with null basis vectors $n_o, n_\infty$ where $n_o\cdot n_\infty = 1$, and $n_o^2 = n_\infty^2 = 0$, as is done in \cite[Table~13.1]{dorst_geometric_2007}.
This approach is desirable because these basis vectors have more geometric meaning, with $n_o$ being associated with the origin, and $n_\infty$ associated with the point at infinity.


We start by defining the conformalized vector space of a real vector space as the triple of (original vector space $V$, $n_0$ coefficient, $n_\infty$ coefficient).
Note that by doing this we \emph{are} choosing a preferred basis (something we generally wanted to avoid) over the extra dimensions, but we continue to avoid doing so over $V$.
\begin{minted}{lean}
@[derive [add_comm_group, vector_space ℝ]]
def conformalize (V : Type*) [inner_product_space ℝ V] : Type* := V × ℝ × ℝ
\end{minted}
We proceed by providing linear maps to extract each component:
\begin{minted}{lean}
def v : conformalize V →ₗ[ℝ] V := linear_map.fst _ _ _
def c_n0 : conformalize V →ₗ[ℝ] ℝ :=
  (linear_map.fst _ _ _).comp (linear_map.snd _ _ _)
def c_ni : conformalize V →ₗ[ℝ] ℝ :=
  (linear_map.snd _ _ _).comp (linear_map.snd _ _ _)
\end{minted}
and some definitions to construct conformal vectors:
\begin{minted}{lean}
/-- The embedding of direction vectors into `conformalize V`. -/
def of_v : V →ₗ[ℝ] conformalize V := linear_map.inl _ _ _
/-- The n₀ basis vector. -/
def n0 : conformalize V := (0, 1, 0)
/-- The n∞ basis vector. -/
def ni : conformalize V := (0, 0, 1)
\end{minted}
Finally, we can define the \emph{up} mapping $\operatorname{up}(v) = n_0 + v + \tfrac{1}{2}{\|v\|}^2n_\infty$ and the conformal metric $Q$, the final pieces needed to construct the Clifford algebra:
\begin{minted}{lean}
/-- The embedding of points from `V` to `conformalize V`. -/
def up (x : V) : conformalize V :=
n0 + of_v x + (1 / 2 * ∥x∥^2 : ℝ) • ni

/-- The metric is the metric of V plus an extra term about n0 and ni. -/
def Q : quadratic_form ℝ (conformalize V) :=
(bilin_form_of_real_inner.comp v v).to_quadratic_form
  - (2 : ℝ) • quadratic_form.lin_mul_lin c_n0 c_ni

variables (V)
/-- Define the Conformal Geometric Algebra over `V` . -/
abbreviation CGA := clifford_algebra (Q : quadratic_form ℝ (conformalize V))
\end{minted}

With our definitions out of the way, our next job is to train the Lean simplifier about trivial combinations of these functions
\begin{minted}{lean}
@[simp] lemma v_of_v (x : V) : (of_v x).v = x := rfl
@[simp] lemma c_n0_of_v (x : V) : (of_v x).c_n0 = 0 := rfl
@[simp] lemma c_ni_of_v (x : V) : (of_v x).c_ni = 0 := rfl
-- 6 more lemmas follow combining {v, c_n0, c_ni} with {n0, ni}
\end{minted}
From this, we can prove that \mintinline{lean}{Q} has the form we'd expect, that \mintinline{lean}{up} correctly produces null vectors, and that the metric between two conformal points is proportional to their distance:
\begin{minted}{lean}
@[simp] lemma Q_apply (x : conformalize V) : Q x = ∥x.v∥^2 - 2 * (x.c_n0 * x.c_ni) :=
by simp [Q, inner_self_eq_norm_sq_to_K]

@[simp] lemma Q_up (x : V) : Q (up x) = 0 :=
by simp [up, ←mul_assoc]

lemma Q_polar_up (x y : V) : quadratic_form.polar Q (up x) (up y) = -dist x y ^ 2 :=
sorry
\end{minted}
Note that while at a glance this last result appears off by a factor of two from the \enquote{usual} result that $\operatorname{up}(x)\cdot\operatorname{up}(y) = -\frac{1}{2}\left\|x - y\right\|$, this is because Lean's \mintinline{lean}{quadratic_form.polar Q x y} is defined as twice the value of the inner product $x \cdot y$.

A construction of PGA in Lean can be done in very much the same way, and is included in our source repository.

\section{Future work}
\label{sec:future}
In \cref{sec:ours}, we formalized a variety of elementary GA operations, and showed how to leverage various existing definitions supplied in \mathlib to do so in a way that is concise and integrates with the rest of the library.
Our work however, is not yet at the point to be ready for use in formalizing results at the edge of research in Clifford Algebras, as some important elementary definitions are still absent.
In this section, we will outline the absent areas of formalization which would contribute significantly to closing this gap.

\subsection{Graded modules and algebras}
To date, \mathlib does not have a reusable definition for graded modules and algebras, making it difficult to make definitions that can be reused across tensor, exterior, and Clifford algebras.
In \cref{sec:ours-z2-grading}, we present a definition that represents the $\mathbb{Z}_2$-gradation using a injective map into \mintinline{lean}{add_monoid_algebra (clifford_algebra Q) (zmod 2)}.
This representation is flawed, as there are terms of this type that do not represent a valid grading result such as \mintinline{lean}{single 1 r} for some scalar \mintinline{lean}{r}, which falsely claims that the scalars are a grade one element.
This can be partially resolved by restricting this injective map to its range to form an equivalence, but the resulting object becomes awkward to work with.

A better solution would be to use \mathlib's \mintinline{lean}{direct_sum} type in place of \mintinline{lean}{add_monoid_algebra}, as this allows us to put a different restriction upon each grade.
Unfortunately, unlike the latter, the former is not equipped with a multiplicative structure needed to use it with the universal property.
Designing the API for equipping such a structure was deemed out of scope for this paper, but will be essential for conveniently working with the grades of geometric, exterior, and tensor algebras in Lean.

\subsection{Isomorphisms from Clifford to exterior algebras}
As mentioned in \cref{sec:ours:wedge}, this work was not able to define the wedge product on multivectors, largely due to desiring a coordinate-free approach.
While this makes our formalization less useful than the other formalizations in \cref{sec:existing-formalizations}, it should be noted that those formalizations require an explicit coordinate parametrization.
To make progress in this direction, \mathlib needs to acquire definitions and API for working with filtered algebras and their associated graded algebra.
Linking up the definitions in \mathlib of the Clifford algebra with the definition of the universal enveloping Lie algebra would help with applying the Poincaré–Birkhoff–Witt theorem, which itself would need formalizing.

\subsection{Geometric Calculus}

Parallel to our efforts to formalize geometric algebra, \mathlib has been making great progress in the field of analysis, with slowly improving support for multivariate calculus.
A notable result in the last few months was the formalization of variations of the Fundamental Theorem of Calculus.

As this part of \mathlib matures, it should be possible to formalize the Geometric Calculus described by \cite{hestenes2012clifford}.
This will however not be a simple task---many of the results are stated in terms of an \mintinline{lean}{is_R_or_C} typeclass, which in practice means they apply only to the reals or the complex numbers.
The next step towards formalizing Geometric Calculus would be to generalize some of these results to the quaternions (which are in \mathlib), which would inform how to sensibly break apart the \mintinline{lean}{is_R_or_C} structure into smaller pieces.

\section{Conclusions}
\label{sec:conclusions}
We have shown how Lean can be used to formalize Geometric Algebra in a variety of ways.
Of these ways, we chose to develop the formalization using a quotient definition, as this offered greater generality than any previous formalization.
We demonstrate how to use the universal property of the Clifford algebra that this definition provides to recover additional definitions and operators, and how to prove that these constructions behave as we intend.
To demonstrate the generality of our approach, we show how to specialize it to a Conformal Geometric Algebra over a vector space of arbitrary dimension.
Throughout, we make liberal use of the \mathlib library for Lean, demonstrating the convenience of having an expansive body of formalized mathematical terminology stated coherently in a single place.

A complete archive consisting of our formalization, our translations of parts of existing formalizations, the precise versions of Lean and \mathlib that our code is compatible with, and a summary of the various contributions to \mathlib made as part of this work can be found on GitHub at
\begin{center}
\url{https://github.com/pygae/lean-ga}
\end{center}
\mathlib is not only growing quickly, but changes quickly too.
Through having contributed the key definitions from our work back into \mathlib, we ensured these definitions will change with it, and can be built upon in future by others.

\section*{Acknowledgements}
We would like to thank the \mathlib community for their valuable and rapid support for all questions about Lean and \mathlib.
The first author is supported by a scholarship from the Cambridge Trust.

\bibliography{references}
\bibliographystyle{spmpsci}


\end{document}